# Radio Interferometers Larger than Earth: Lessons Learned and Forward Look of Space VLBI

**Leonid I. Gurvits[a,b]**

[a] *Joint Institute for VLBI ERIC (JIVE), Oude Hoogeveensedijk 4, 7991 PD Dwingeloo, The Netherlands*
[b] *Dept. of Astrodynamics and Space Missions, TU Delft, Kluyverweg 1, 2629 HS Delft, The Netherlands*
  lgurvits@jive.eu

**Abstract**

Extension of radio interferometric baselines into space is inevitable if a diffraction-limited angular resolution defined by the Earth diameter at a given observing wavelength limits a pursuit of specific scientific goals. This was understood in the early1960s, at the very dawn of the era of Earth-based Very Long Baseline Interferometry (VLBI). Since then, three VLBI missions operated in Space thus enabling baselines longer than the Earth diameter. These are the Tracking and Data Relay Satellite Orbital VLBI experiment (1986–1988), the VLBI Space Observatory Program (VSOP, 1997–2003) and RadioAstron (2011– present time). These Space VLBI (SVLBI) systems enabled studies of celestial radio sources with an unprecedentedly sharp angular resolution reaching 0.1 nanoradian (tens of microarcseconds) and sharper. Space VLBI is a conglomerate of diverse technologies and arguably one of the most sophisticated techniques of space-based astronomy. It involves deployment of large space-borne electro-mechanical structures, state of the art analogue and digital instrumentation, precise attitude control and orbit determination, high capacity data downlink, simultaneous and highly coordinated operations of space-borne and Earth-based facilities, advanced processing of large (tens of terabytes) volumes of data. The first generation SVLBI missions provided cutting-edge results in several topics of modern radio astronomy. These include discoveries of ultra-compact galactic hydroxyl and water vapor masers, radio emission in active galactic nuclei with the brightness exceeding conventional theoretical limits, and detection of pulsar emission at meter wavelengths shedding a new light on the properties of the interstellar medium. One of the Space VLBI missions, the RadioAstron, also ventured into the domain of fundamental physics by sing its onboard Hydrogen maser local oscillators for experimental verification of the Einstein Equivalence Principle. The paper provides a brief review these discoveries and lessons learned over the first half a century of Space VLBI. This comparison, together with the review of technological achievements of the three SLBI missions, creates a fundament for projecting the development of space-based radio interferometry into the next decades.

## 1. Introduction

Radio astronomy, born in 1933 [1], exemplifies the multi-wavelength nature of the modern astrophysics. During the first two decades of its development, radio astronomy observations were conducted with antennas limited in their angular resolution ("sharpness of vision"). In the case of reflecting antennas (in a widely used professional slang – the "dishes"), the resolution is defined by the diffraction limit, $\lambda/D$, where $\lambda$ is the wavelength, and $D$ – diameter of the reflector, just as is the case in the "traditional" optical astronomy. For typical in radio domain decimeter to meter wavelengths, affordable dishes of tens of meters in diameter can reach angular resolution of tens of arcminutes, far worse than a typical angular resolution of Earth-based optical telescopes, which is of the order of an arcsecond. A concept of interferometers was introduced in radio astronomy in the beginning of the 1950s [2]. It enabled radio astronomers to sharpen the angular resolution defined by the same simple formula mentioned above, with substitution of the antenna diameter to the projection of the baseline, connecting two elements of the interferometer, on the picture plane. Later, in the second half of the 1960s, the concept of radio interferometry got its ultimate extension to the baselines, comparable to the Earth diameter. This was the beginning of Very Long baseline Interferometry (VLBI) [3]. VLBI offered a breathtaking sharpness of vision, reaching milliarcseconds and even sub-milliarcseconds. However, very soon radio astronomers realized that even this record-high angular resolution, that exceeded best optical values by three orders of magnitude, was not sufficient: there were celestial radio sources that remained point-like (unresolved) even in observations with the global VLBI systems. In fact, the necessity of even higher angular resolution was understood at the very dawn of the era of Earth-based Very Long Baseline Interferometry (VLBI). As is clear from the simple formula at the beginning of this introduction, the only solution of getting an even sharper resolution at a given wavelength than offered with a baseline equal to the Earth diameter is to place at least one element of an interferometer in Space. This is Space VLBI (SVLBI).

## 2. SVLBI proof of concept – TDRSS experiment

The first ad hoc demonstration of SVLBI experiment was conducted with the NASA's geostationary Tracking and Data Relay Satellite (TDRS) in 1986 [4]. The satellite is shown in Fig. 1. It was equipped with two 4.9 m antennas, one of which was used as a radio telescope. The observations were conducted at 13 and 2 cm together with a network of large Earth-based antennas in Australia, Japan and the US.

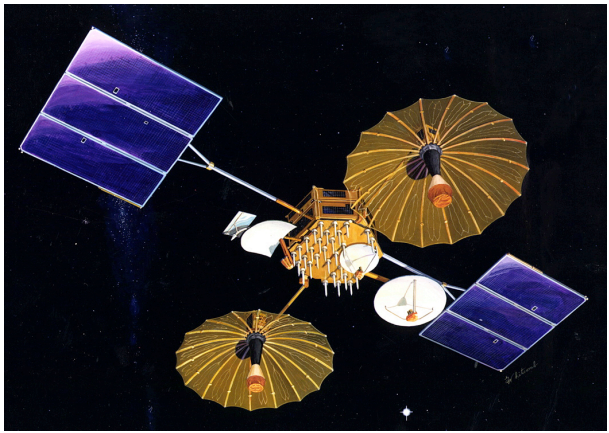

Fig. 1. TDRSS communication satellite in orbit, an artist's impression. Credit: NASA.

The space-borne component of this first SVLBI system was not designed as a radio telescope, and was not supposed to operate as such. It was the ingenuity of the NASA JPL engineers who were able to re-configure and re-program the on-board instrumentation in the way making it compatible with the Earth-based VLBI radio telescope.

The main achievement of the TDRSS Orbital VLBI experiment was the confirmation of a principle possibility of obtaining an interferometric response (so called "interferometric fringes") on baselines exceeding the Earth diameter and involving a radio telescope not fixed on the surface of Earth. The maximum projected baseline achieved in the demonstration was equal to 2.2 Earth diameters (ED). It has to be noted, that the sensitivity of an interferometer to the source's brightness, traditionally measured in radio astronomy in degrees of Kelvin (of a hypothetic black body of the same brightness as the source), depends on the physical length of the projected baseline. In other words, the source can be resolved with a given baseline only up to a certain brightness temperature. If the brightness is higher for a given total flux density of the source, a longer baseline is needed for resolving the source's image. The TDRSS Orbital VLBI experiment indicated that in several quasars the brightness temperature is higher than can be measured with Earth-based VLBI systems.

## 2. The first generation SVLBI missions: VSOP and RadioAstron

In the wake of the success of the TDRSS VLBI demonstration experiment, the VLBI Space Observatory Program (VSOP) was conceived at the Institute of Space and Astronautical Science in Japan in the end of the 1980s [5]. The concept presumed a launch on the new four-stage solid fuel rocket M-V and use as the mission platform the engineering test satellite Muses-B. Although the project was formally aimed primarily at the test of the new rocket and new satellite platform, it did have a core science program that defined the science payload.

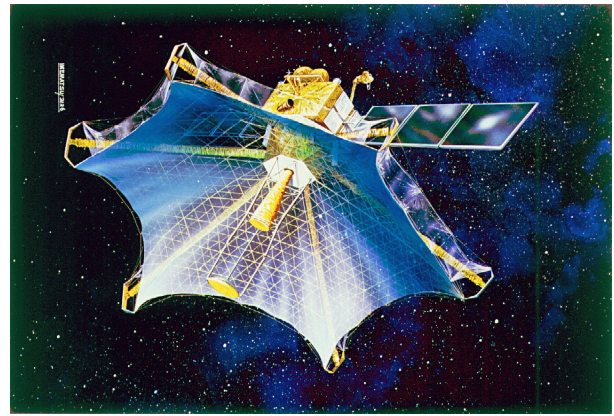

Fig. 2. VSOP HALCA satellite in orbit, an artist's impression. Credit: JAXA.

The main component of the mission was the space-borne radio telescope with the parabolic reflector of 8.8 m effective diameter (Fig. 2). Its surface was formed by a gold-coated molybdenum wire mesh. Three receivers enabled observations at 18, 6 and 1.3 cm in left-hand circular polarisation (LCP). The satellite was on a medium eccentricity orbit with the apogee of about 22,000 km above Earth and orbital period of about 6 hrs. The mission had a sizeable involvement of international community coordinated by the group called VISC (VSOP International Science Council). Among other things, VISC coordinated such the mission implementation issues as operations of five dedicated Earth-based tracking and data acquisition stations (in Australia, Japan and Spain, one in each country, plus two stations in the US), orbit determination, VLBI data processing, Earth-based VLBI network arrangements and policy of access to the mission by the worldwide science community. After successful launch on 12 February 1997, the satellite was renamed HALCA (Highly Advanced Laboratory for Communication and Astronomy).

VSOP operated with the aggregate data rate of 128 Mbit/s – the highest sustainable data downlink rate of a space science mission at the time. The signal coherency – the major requirement of a VLBI system – was supported by a phase-locked loop (PLL), fed by a hydrogen maser frequency standard at a tracking station. Two of the three observing bands, 18 and 6 cm, operated successfully. The signal in the third, the highest frequency band of 1.35 cm, had a very high system noise, preventing science operations. It is believed that the reason was a damage of the waveguide of this band due to a high level of vibration during the satellite launch. Three data processing (correlation) facilities supported the mission. They were located at the Dominion Radio Astrophysical Observatory in Penticton, Canada; National Astronomical Observatory of Japan in Mitaka, Japan; and National Radio Astronomy Observatory in Socorro, NM, USA. The HALCA satellite operated for six years until 2003 and was switched off after its attitude control reaction wheels stopped operating after exceeding their warranted period significantly.

The science program of the VSOP included a major mission-led continuum VLBI Survey of Active Galactic Nuclei [6], which had taken about 25% of the mission operational time. About 50% of the operational time was open to the worldwide scientific community on the basis of peer-reviewed science proposals without any pre-assigned time quotas or guaranteed observing time (the so called "open sky" access). The remaining 25 of the operational time were taken by tests and other engineering operations.

The VSOP HALCA mission provided a wealth of observational material on continuum (both 18 and 6 cm wavelengths) and hydroxyl spectral line observations at 18 cm on compact structures of galactic and extragalactic radio sources at the hitherto unchartered sub-milliarcsecond angular scale. The mission was highly acclaimed: it was awarded with the Team Achievement award of the International Academy of Astronautics in 2005.

The next and so far the only other first-generation SVLBI mission is the Russian-led RadioAstron [7]. The project was conceived in the late 1970s at the Space Research Institute of the USSR Academy of Sciences and was launched from Baikonur on a Zenit-2SB rocket and a Fregat booster stage on 18 July 2011. The Astro Space Center of the Lebedev Physical Institute of the Russian Academy of Sciences and the Lavochkin Science and Production Association lead the mission. A modification of the universal satellite platform Navigator, called Spektr-R, carries the space-borne 10 m radio telescope (Fig. 3).

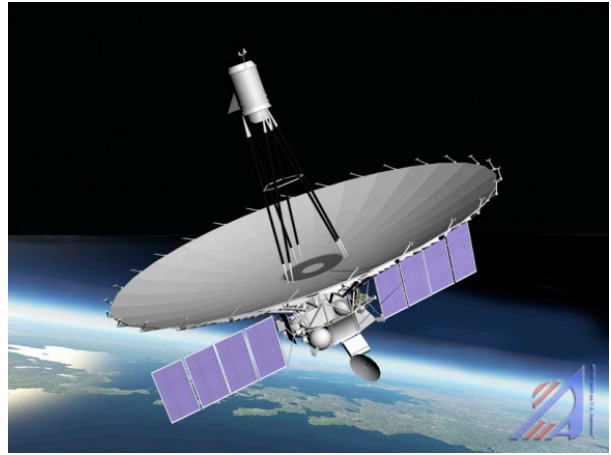

Fig. 3. RadioAstron spacecraft Spektr-R in orbit, an artist's impression. Credit: Lavochkin Association.

The main component of the RadioAstron's science payload is the 10-m parabolic reflector formed by 27 carbon-fiber petals and a central circular part. The telescope is equipped with four radio astronomy dual-polarisation (both left- and right-hand, LCP and RCP) for the wavelengths 92, 18, 6 and 1.3 cm. The satellite is on a highly eccentric evolving orbit with its apogee reaching the distance of 340,000 km from Earth (~28 ED) with the orbital period of about 9 days. Just as VSOP, RadioAstron is able to use a highly stable heterodyne signal of an Earth-based hydrogen maser frequency standard through a PLL. However, the RadioAstron spacecraft carries also two active Hydrogen maser standards. One of these devices has been used as the prime provider of the highly stable heterodyne signal for 6 years, twice longer than the warranted period, from the launch until July 2017. Just as in the case of VSOP, RadioAstron has a large involvement of international community in its mission operational and science components. Similarly to VSOP, the RadioAstron International Science Council (RISC) coordinates the global international involvement in the mission. The Earth-based support is provided by two data acquisition stations in Pushchino (Russia) and Green Bank (WVA, USA). They receive the raw VLBI data stream from the satellite with the data rate of 128 Mbit/s in the same data format as VSOP. The mission OD is based on measurements provided by numerous radio tracking stations of the Roscosmos in Russia as well as the global international network of laser tracking stations. The RadioAstron VLBI data are processed at three correlators operated at the Astro Space Center in Moscow (Russia), Max-Plank-Institut für Radioastronomie in Bonn (Germany) and Joint Institute for VLBI ERIC in Dwingeloo (The Netherlands).

After completion of the in-orbit checkout and verification, the mission has become open to the worldwide science community under the "open sky" access policy. Since 2012, RadioAstron has issued 6 Calls for science proposals, each covering one operational year. Due to the achievable baselines reaching 28ED, the RadioAstron mission has established angular resolution records in all four operational bands, and the absolute record of about 7 microarcseconds at the shortest observing wavelength of 1.3 cm.

The two first-generation SVLBI missions, VSOP and RadioAstron, which followed the first demonstration SVLBI experiment with TDRSS, created a solid technological basis for further development of space-borne radio interferometry. Table 1 summarizes their major technical parameters.

Table 1. Major parameters of the first three operational SVLBI systems

|  | TDRSS | VSOP | RadioAstron |
|---|---|---|---|
| In orbit ops. | 1986-88 | 1997-03 | 2011–pt |
| Antenna [m] | 5.8 | 8.8 | 10 |
| $B_{max}$ [ED]* | 2.2 | 3 | 27 |
| Wavelength [cm] | 13, 2 | 18, 6 | 92, 18, 6, 1.3 |
| Data rate [Mbps] | 28 | 128 | 128 |

* Maximal baseline projection in Earth Diameters (ED)

**3. Science legacy of the first generation SVLBI**

In a broad sense, one might expect that radio astronomy tools operating at the highest angular resolution should address three major science topics.

First, the emission mechanism of continuum sources, primarily extragalactic sources associated with Active Galactic Nuclei (AGN) powered by supermassive black holes, should be determined. The existing paradigm of AGN considers the synchrotron mechanism as the main mechanism of their electromagnetic emission. However, important details of these extremely powerful emitters remain enigmatic. The key to the understanding of the physics of these "most powerful engines" of the Universe lies in the close vicinity to their central areas, parsecs and sub-parsecs in linear size. To zoom into these objects, located at the distances of mega- and gigaparsecs from the Earth, one needs to achieve angular resolution going down to microarcseconds. One particularly daring question is related to the highest brightness temperature in these sources. As it has been demonstrated by Kellerman and Pauliny-Toth in 1969 [8], the inverse Compton scattering limits the brightness temperature of steady synchrotron emission by the value of $10^{11.5}$ K. An even stricter limit of $10^{11.5}$ K has been formulated by Readhead in 1994 [9] based on the assumption of equipartition of energy density in AGN cores between particles and magnetic field. But do these limits hold in reality, thus confirming the current understanding of the physics of AGN? To answer this question one must observe the phenomenon with baselines longer than the Earth diameter.

The second major drive for the highest angular resolution in radio astronomy comes from studies of ultra-compact sources of narrow-band spectral line emission of cosmic masers (see [10] and references therein). At present, most "popular" molecules, responsible for maser emission are water vapor $H_2O$, emitting at 1.35 cm, and hydroxyl OH, emitting at 18 cm, but other molecules and spectral lines are also queuing for detailed high-resolution studies. For many of these molecules and sources, both galactic and extragalactic, the Earth-based VLBI studies are limited in resolution, insufficient for detailed zoom into the areas of maser emission generation.

The third area of attention for high-resolution radio astronomy is the impact of propagation media – intergalactic, interstellar, interplanetary, and the media in the immediate vicinity of Erath, its ionosphere and atmosphere. Some of the listed media are important subjects of astrophysical interest; all act as agents, interfering in the attempts to get clear and sharp picture of remote sources of radio emission. An attempt to predict the impact of propagation effects on Space VLBI has been exercised in 1990, during the transition from the first demonstration of SVLBI by the TDRSS Orbital VLBI experiment and the first two dedicated SVLBI missions, RadioAstron and VSOP [11].

The science legacy of the VSOP mission is presented in details in two books [12, 13]. In brief, the VSOP science accomplishments can be presented by the following (subjective) list:
- Indication on the brightness temperature in AGN exceeding the Inverse Compton limit, achieved in the VSOP AGN Survey [14] and observations of the brightest blazars, e.g. AO 025+164 [15].
- Detection of unexpectedly compact structures in stellar OH masers [16].
- Detailed high-resolution high-dynamic range images of several strong AGN (see [12] and references therein].

At the time of this writing, the RadioAstron mission is still operating. So, it is premature to formulate its science legacy. However, several RadioAstron results already create a solid basis for the mission's legacy. Not pretending for completeness, the list of the RadioAstron major results include:
- Direct measurements of brightness temperatures

in AGN, exceeding the conventional Inverse Compton scattering limit (see [17] and results of the ongoing RadioAstron AGN Survey project, in preparation). This result requires a critical look into the conventional physical model of AGN electromagnetic emission.
- Detection of refractive substructure in Sgr A* [18], the quasar 3C 273 [19], and pulsars [20].
- Detection of compact maser spots in galactic and extragalactic sources [21]

In addition, the presence of the H-maser frequency standard on board the Spektr-R satellite made it possible an ad hoc gravitational redshift experiment. Although not planned at the stage of the mission development and therefore not involved in the formation of the requirements for the mission science payload, this experiment enables to verify the Einstein Equivalence Principle at the level no worse than the dedicated mission Gravitational Probe A, conducted in 1978, and possible better [22].

**4. Lessons learned (so far) and future Space VLBI**

Over the past four decades no less than two dozens of more or less detailed proposals and design studies of Space VLBI missions have been conducted around the world. To date, only three of them briefly described above have reached the orbit. They created the basis, which would serve all future missions in the high-resolution domain of radio astronomy.

Extensive analysis of the first generation Space VLBI is underway and goes well beyond the scope of this review. However, as a preview of future detailed analysis, several mission characteristics that limit the science productivity of SVLBI should be mentioned. In an arbitrary order they are:
- Limited sensitivity in both continuum and spectral line regimes due to a small size of the space-borne radio telescopes (at least an order of magnitude smaller in collecting area than typical Earth-based VLBI radio telescopes).
- Limited sensitivity in continuum observations due to the data downlink "bottleneck", 128 Mbit/s for both VSOP and RadioAstron, comparing to routinely available in modern Earth-based VLBI systems data rates of 1 Gbit/s and higher.
- An underestimate of operational impact of various functional limitations of SVLBI systems, especially the space-borne radio telescopes, on the science return and overall efficiency of the missions. An attempt to evaluate just attitude limitations at the design phase of the RadioAstron mission, however severe it looked at that time [23], still proved to be too optimistic. In reality, other constraints, such as thermal and availability of Earth-based co-observing telescopes, limit the overall mission efficiency severely.

These and other mission characteristic should be a subject of design analysis of prospective SVLBI missions. The major will aim at an increased sensitivity, wider frequency domain coverage, higher degree of observing efficiency and operational versatility. The following four areas of developments appear to be most topical from the today's standpoint.

*1.1 Space–Space baselines*

All three implemented to date SVLBI systems had only one space-borne component and exploited networks of Earth-based telescopes for achieving as a complete coverage of spatial frequencies (the so called *uv*-coverage) as feasible or achievable. However, space-space baselines, in the case of multiple space-borne SVLBI components add two new major qualities: (i) they might offer even longer baselines than those to Earth-based telescopes; (ii) they would enable faster and more complete *uv*-coverage; (iii) they are free of the impact of the Earth atmosphere and ionosphere at the Earth-based telescopes' end; and (iv) they are free of visibility limitations defined by the Earth's horizon at each Earth-based telescope site. Indeed, several design studies have suggested capitalising on these advantages of multiple (at least two) SVLBI spacecraft. These are the iARISE project [24] at cm-mm wavelengths, the Chinese SVLBI projects at millimetre [25] and decimetre-meter [26] wavelengths. Recently, a concept of a Space-to-Space interferometer for high-quality *uv*-sampling with two or more space-borne radio telescopes on Medium Earth Orbits has been proposed [27].

*1.2 Millimetre and sub-millimetre wavelengths*

Earth-based mm-VLBI is limited by the opaqueness of the Earth atmosphere. The most advanced mm-VLBI Earth based facilities, such as the coordinated Global VLBI Array (GMVA, [28]) and Event Horizon Telescope [29], operate at wavelengths no shorter than 1.5 mm. However, the science case for VLBI at shorter millimetre and sub-millimetre wavelengths (frequencies reaching terahertz and higher) is very strong. In the past two decades, several design studies addressed prospective mm-SVLBI facilities, such as ARISE [30] and VSOP-2 [31], but proved to be unaffordable. A follow-up on the RadioAstron mission, a mm-wavelength mission Millimetron is being under development in Russia [32]. A new impetus for assessing sub-mm SVLBI is triggered be the tantalising perspective of having an unobscured view at the shadow of supermassive black holes in galactic nuclei – a potentially key experiment in fundamental physics. Indeed, prospects of such the advanced sub-mm SVLBI system have been recently addressed at several workshops and studies [33].

*1.3 Ultra-long wavelengths*

At the opposite to millimetre domain of the radio part of the electromagnetic spectrum, an interest for radio astronomy studies is growing too. It requires to venture into the hitherto unexplored (in fact, the last unexplored) region of the electromagnetic spectrum, wavelengths longer than ~15 m, the so-called ultra-long-wavelengths (ULW). At these wavelengths, the Earth's ionosphere is opaque and prevents cosmic radio emission from reaching Earth-based radio telescopes. The only opportunity of observing cosmic radio emission in the ULW regime is to place a radio telescope in space. While "single-dish" observations in Space would be a valuable start of the era of ULW astronomy, the most attractive seems to be space-borne ULW radio interferometry. Its science outlook covers a broad range of disciplines from cosmology to extragalactic and galactic astrophysics to solar and planetary science to geophysics. A number of initiatives and design studies address the ULW astronomy perspectives, especially in interferometric configurations (see [34], [35] and references therein).

*1.4 SVLBI and astrometry*

A highly attractive application of sharpest resolution of Space VLBI might emerge in ultra-precise astrometry. The first generation SVLBI systems were unable to conduct astrometric VLBI measurements due to the insufficiently accurate orbit determination and inability to conduct typical for astrometric VLBI "nodding" and phase-referencing observations. Such the ability was a mission specification requirement for ARISE and VSOP-2. Such the requirement will be reinstated in the design characteristics of the future SVLBI systems, especially those operating at shorter wavelengths (thus, with a higher angular resolution) and able to provide rapid response to transients, including aftermaths of gravitational-wave generating events.

## 5. Conclusions

The science and technology challenges briefly described in this review present SVLBI as a complicated and costly science endeavour. However, the TDRSS Orbiting VLBI experiment and two first-generation SVLBI missions, VSOP and RadioAstron, demonstrate the feasibility and high scientific potential of high-resolution Space VLBI studies. While the science goals listed in Section 4 appear to be extremely attractive from the today's standpoint, perhaps the most valuable outcome of future Space VLBI studies will come in the area hardly anticipated today – just because next generation SVLBI systems will open up a new large area of parameter space, extremely small angular scales. As was always the case in history of astronomy, such breakthroughs in new characteristics of observing systems never betray in providing new science insight.

**Acknowledgements**

The VSOP Project was led by the Institute of Space and Astronautical Science (Japan) in cooperation with many agencies, institutes, and observatories around the world.VSOP, The RadioAstron project is led by the Astro Space Center of the Lebedev Physical Institute of the Russian Academy of Sciences and the Lavochkin Association of the Roscosmos State Corporation for Space Activities, and is a collaboration with partner institutions in Russia and other countries.